\documentclass[10pt]{iopart2}

\usepackage{amsmath,amssymb}
\usepackage{iopams}
\usepackage{bm,mathrsfs,color}
\usepackage{graphicx}
\usepackage{braket}

\newcommand{\yf}[1]{{#1}}

\begin{document}

\title[Large longitudinal magnetoresistance of multivalley systems]
{Large longitudinal magnetoresistance of multivalley systems}

\author{Yuki Mitani$^1$ \& Yuki Fuseya$^{1, 2}$}

\address{
$^1$ Department of Engineering Science, University of Electro-Communications, Chofu, Tokyo 182-8585, Japan}
\address{$^2$ Institute for Advanced Science, University of Electro-Communications, Chofu, Tokyo 182-8585, Japan}
\ead{mitani@kookai.pc.uec.ac.jp}
\vspace{10pt}

\begin{abstract}
	The longitudinal magnetoresistance (MR) is assumed to be hardly realized as the Lorentz force does not work on electrons when the magnetic field is parallel to the current. However, in some cases, longitudinal MR becomes large, which exceeds the transverse MR. To solve this problem, we have investigated the longitudinal MR considering multivalley contributions based on the classical MR theory. We have showed that the large longitudinal MR is caused by off-diagonal components of a mobility tensor. Our theoretical results agree with the experiments of large longitudinal MR in IV-VI semiconductors, especially in PbTe, for a wide range of temperatures, except for linear MR at low temperatures.
\end{abstract}

%
\vspace{2pc}
\yf{\noindent{\it Keywords}: magnetoresistance, large longitudinal  magnetoresistance, multivalley systems, PbTe}
%
%
\ioptwocol

\section{Introduction}
Magnetoresistance (MR) is one of the most fundamental phenomena in solid-state physics. It was discovered by Thomson (Lord Kelvin) in 1857 \cite{Thomson1857} and is available in every standard textbook of solid-state physics \cite{Ziman_book,Kittel_book_quantum,Abrikosov_book}. Some monographs specialized in MR are also available \cite{Beer1963,Pippard_book}. It seems that the physics of MR is old and well understood. However, many factors have not reached a consensus, e.g., (quasi) linear MR \cite{Kapitza1928,Taub1971,RXu1997,Narayanan2015,ZZhu2018,Kawakatsu2019} and negative longitudinal MR \cite{Arnold2016,JXu2019}.
The large longitudinal MR is one of such fundamental mysteries that need to be explored. According to the textbook knowledge, the transverse MR arises for semimetals or multivalley systems \cite{Ziman_book,Kittel_book_quantum,Abrikosov_book,ZZhu2018}. On the other hand, the longitudinal MR never arises. However, the longitudinal MR often realizes experimentally. One striking example is the large longitudinal MR in IV-VI semiconductors \cite{Allgaier1958,Gupta1978,Onuki_private}.

Allgaier detailed the transverse and longitudinal MR of IV-VI semiconductors (PbS, PbSe, and PbTe) at 295, 77.4, and 4.2 K for various n- and p-type samples. Each sample exhibits anomalously large longitudinal MR, which is comparable to transverse MR. Especially, in PbTe, the longitudinal MR is larger than the transverse MR. Although the large longitudinal MR in PbTe has been reported repeatedly \cite{Gupta1978,Onuki_private}, its mechanism has not been clarified yet. The electronic structure of IV-VI semiconductors is quite different from group IV semiconductors or III-V semiconductors: there are four ellipsoidal Fermi surfaces of doped carriers at the $L$ points in the Brillouin zone but not at the $\Gamma$ point; i.e., IV-VI semiconductors are being referred to as multivalley systems.

In this paper, we theoretically analyze the longitudinal and transverse MR of multivalley systems. 
Although some theoretical studies showed that the longitudinal MR can be realized in multivalley systems \cite{Abeles1954,Shibuya1954,Gold1957,Roth1992,Askerov1994} but its mechanism has not been clarified yet, which is the main subject of this paper.
\yf{Especially, we have not reached an intuitive explanation of why the large longitudinal MR is realized. In addition, the quantitative explanation for the experimental results on PbTe has not been given yet, which is the second subject of this paper.}
We adopt the MR theory in the tensor form, which makes the calculation very transparant, resulting in the clarification of the mechanism of large longitudinal MR. The tensor form of MR theory was derived by Mackey and Sybert based on the Boltzmann equation \cite{Mackey1969}, while  here we reformulate their theory based on the classical equation of motion.

\section{Theory}
The classical equation of motion in external electric ($\bm{E}$) and magnetic ($\bm{B}$) fields is given as
\begin{align}
	m^* \frac{d\bm{v}}{dt}= \pm e (\bm{E} + \bm{v}\times \bm{B}) -\frac{m^*}{\tau}\bm{v},
\end{align}
where $m^*$ is the effective mass, $\bm{v}$ is the velocity, and $\tau$ is the relaxation time of carriers. $e>0$ is the elementary charge, and the upper (or lower) sign corresponds to the charge of the holes (electrons). In the static limit ($d\bm{v}/dt =0$), the current density $\bm{j}= \pm n e\bm{v}$ becomes
\begin{align}
	\bm{j}=  ne \hat{\mu}\cdot \left(\bm{E}\pm \frac{1}{ne}\bm{j}\cdot \hat{B} \right).
\end{align}
In this study, we have introduced the mobility tensor $\hat{\mu}$ and the magnetic tensor $\hat{B}$:
\begin{align}
	\hat{\mu}=\begin{pmatrix}
		\mu_{xx} & \mu_{xy} & \mu_{xz}
		\\
		\mu_{yx} & \mu_{yy} & \mu_{yz}
		\\
		\mu_{zx} & \mu_{zy} & \mu_{zz}
	\end{pmatrix}
	,  \quad
	\hat{B}=\begin{pmatrix}
		0 & -B_z  &B_y \\
		B_z & 0 & -B_x \\
		-B_y & B_x & 0
	\end{pmatrix}.
\end{align}
\yf{The mobility tensor can be expressed in terms of the effective mass tensor $\hat{m}$ as $\hat{\mu}=e \tau \hat{m}^{-1}$, if one adopt the constant relaxation time $\tau$ approximation.}
\yf{The magnetic tensor has the same form as the electromagnetic tensor that appears in special relativity. It expresses the outer product as $\bm{j}\cdot \hat{B}$ instead of $\bm{j}\times \bm{B}$.}
The conductivity tensor $\hat{\sigma}$, which is defiend by $\bm{j}=\hat{\sigma}\cdot \bm{E}$, is then obtained as
\begin{align}
	\hat{\sigma}=ne \left( \hat{\mu}^{-1} \pm \hat{B} \right)^{-1}.
	\label{Formula}
\end{align}
This is equivalent to the Mackey--Sybert’s formula based on the Boltzmann theory \cite{Mackey1969,Awashima2019}, though Eq. \eqref{Formula} is obtained from the classical equation of motion. For the multivalley systems, the total conductivity tensor is obtained by summing up the contributions of each valley $i$, and the total magnetoresistivity, $\hat{\rho}$, is given by \cite{Aubrey1971,Collaudin2015}
\begin{align}
	\hat{\rho}&=\left( \sum_i \hat{\sigma}_i \right)^{-1},
	\label{rho}
	\\
	\hat{\sigma}_i &=n_i e \left( \hat{\mu}_i^{-1} \pm \hat{B} \right)^{-1}.
	\label{sigmai}
\end{align}
There is no limit for the magnetic field strength in the above derivation.
\yf{The carrier density can be different among the valleys at high enough magnetic field (in the quantum limit). In this paper, we assume that each valley has the equivalent carrier density, i.e., $n=n_i$ hereafter.}

\subsection{Two-valley system}
\begin{figure}
	\begin{center}
		\includegraphics[width=8cm]{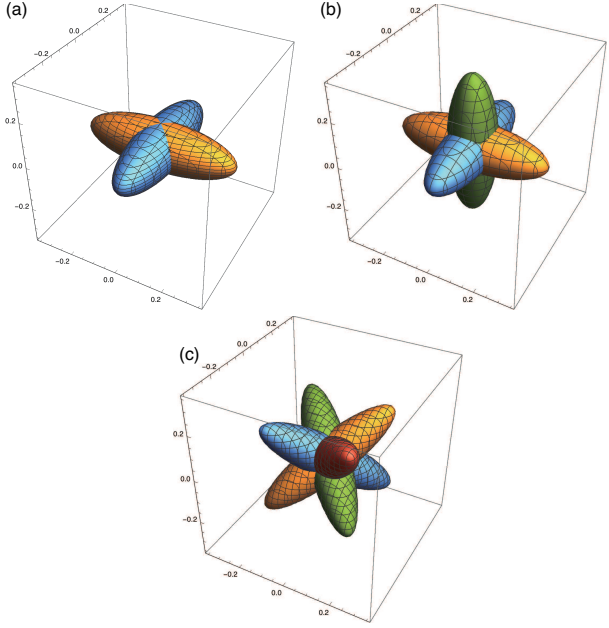}
		\caption{\label{Fig1} Fermi surfaces of (a) two-valley, (b) three-valley, and (c) four-valley models. In general, the valleys are located at different points in the Brillouin zone; in this figure, they are displayed together at the center.
		}
	\end{center}
\end{figure}
First, we analyze the simplest case of the multivalley system, i.e., two-valley system, in which one ellipsoidal Fermi surface is orthogonal to another [Fig. \ref{Fig1} (a)]. The mobility tensors are expressed as
\begin{align}
	\hat{\mu}_1= \mu_0 \begin{pmatrix}
		\beta & 0 & 0 \\
		0 & 1 & 0\\
		0 & 0 & 1
	\end{pmatrix},
	\quad
	\hat{\mu}_2= \mu_0 \begin{pmatrix}
		1 & 0 & 0 \\
		0 & \beta & 0\\
		0 & 0 & 1
	\end{pmatrix},
	\label{mu12}
\end{align}
where $\mu_0$ is the constant mobility \yf{and $0<\beta <1$}. In this model, $(1-\beta)^2$ denotes the anisotropy of the mobility.

\begin{figure}
	\begin{center}
		\includegraphics[width=7cm]{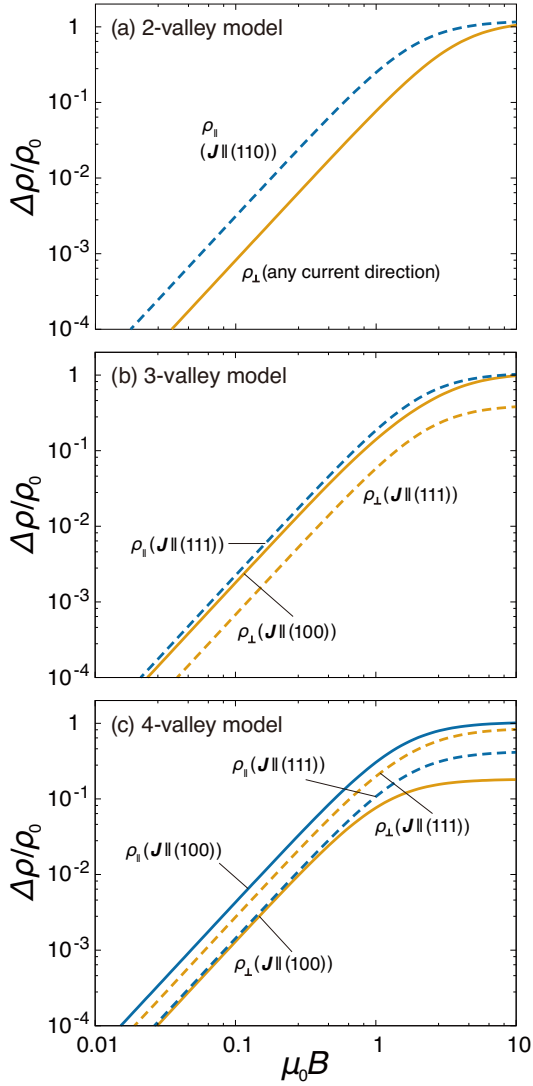}
		\caption{\label{Fig7} Magnetic field dependence of $\Delta \rho_{\perp, \parallel}$ regarding (a) two-valley ($\beta=0.15$), (b) three-valley ($\beta=0.15$), and (c) four-valley models  ($\gamma=0.39$).
		}
	\end{center}
\end{figure}

\yf{The conductivities of each valley are obtained as
\begin{subequations}
\begin{align}
		\sigma_{1xx}&=
		ne\mu_0\beta\left(\mu_0^2B_x^2+1\right)/A_1, \\
	\sigma_{1yy}&=
		ne\mu_0\left(\beta\mu_0^2B_y^2 + 1\right)/A_1,\\
	\sigma_{1zz}&=
		ne\mu_0\left(\beta\mu_0^2B_z^2 + 1\right)/A_1,\\
	\sigma_{1xy}&=
		ne\mu_0\beta\left(\mu_0^2B_xB_y+\mu_0B_z\right)/A_1, \\
	\sigma_{1yz}&=
		ne\mu_0\left(\beta\mu_0^2B_yB_z + \mu_0B_x\right)/A_1, \\
	\sigma_{1zx}&=
		ne\mu_0\beta\left(\mu_0^2B_zB_x-\mu_0B_y\right)/A_1,
\end{align}
\label{sigma1}
\end{subequations}
\begin{subequations}
\begin{align}
	\sigma_{2xx}&=
		ne\mu_0\left(\beta\mu_0^2B_x^2+1\right)/A_2, \\
	\sigma_{2yy}&=
		ne\mu_0\beta\left(\mu_0^2B_y^2+1\right)/A_2,\\
	\sigma_{2zz}&=
		ne\mu_0\left(\beta\mu_0^2B_z^2 + 1\right)/A_2,\\
	\sigma_{2xy}&=
		ne\mu_0\beta\left(\mu_0^2B_xB_y+\mu_0B_z\right)/A_2, \\
	\sigma_{2yz}&=
		ne\mu_0\beta\left(\mu_0^2B_yB_z+\mu_0B_x\right)/A_2, \\
	\sigma_{2zx}&=
		ne\mu_0\left(\beta\mu_0^2B_zB_x + \mu_0B_y\right)/A_2,
\end{align}
	\label{sigma2}
\end{subequations}
where $A_1 = 1+\mu_0^2\left[ B_x^2 +\beta\left(B_y^2+B_z^2\right) \right]$ and $A_2 = 1+\mu_0^2\left[ B_y^2 +\beta\left(B_x^2+B_z^2\right) \right]$. The other off-diagonal elements can be obtained by the relation $\sigma_{\mu \nu}(\bm{B})=\sigma_{\nu \mu}(-\bm{B})$.
}
\yf{First, we} fix the current direction as $\bm{J}\parallel (1 0 0)$. The transverse MR, $\bm{B}= (0, 0, B)$, and the longitudinal MR, $\bm{B} = (B, 0, 0 )$, are obtained from Eqs. \eqref{rho}, \eqref{sigmai}, \yf{\eqref{sigma1}, and \eqref{sigma2}} in the following forms:
\begin{align}
	\frac{\Delta \rho_{\perp}}{\rho_0} &= \frac{\beta (1-\beta)^2 (\mu_0 B)^2 }{(1+\beta)^2 +4\beta^2 (\mu_0 B)^2  },
	\label{TMR1}
	\\
	\frac{\Delta \rho_\parallel}{\rho_0}&=0,
	\label{LMR1}
\end{align}
where $\Delta \rho_{\perp, \parallel} = \rho_{\perp, \parallel}-\rho_0$, and $\rho_0 = 1/[ne\mu_0(1+\beta)]$ is the resistivity at zero field.
\yf{$\rho_{\perp}$ corresponds to $\rho_{xx}$ in $\bm{B}=(0, 0, B)$ and $\rho_{\parallel}$ corresponds to $\rho_{xx}$ in $\bm{B}=(B, 0, 0)$.
}
The MR is usually written as a function of $\mu_0 B=\omega_c \tau$, where $\omega_c = eB/m^*$ is the cyclotron frequency. The longitudinal MR never arises along this direction, whereas the transverse MR arises as it is proportional to the anisotropy $(1-\beta)^2$. $\Delta \rho_\perp \propto B^2$ at low field intensities, $\mu_0 B \lesssim 1$, and saturates at strong field limits, $\mu_0 B \gg 1$, as shown by the solid line in Fig. \ref{Fig7} (a).
\yf{
(Here, we set $\beta=0.15$ to be consistent with the experiments on PbTe as discussed later. The qualitative behavior of $\Delta \rho_{\perp, \parallel}$ is unchanged if we change $\beta$.)
}
These results for multivalley systems are well known \cite{ZZhu2018}.

Next, we consider a case by rotating the {\it current} orientation. We rotate the valley and fix the current orientation instead of rotating the current orientation and fixing the mobility. With respect to the rotation in the $x$--$y$ plane, the mobility tensors are given by $\hat{\mu}_i (\theta)=\hat{R}^{-1}(\theta)\hat{\mu}_i\hat{R} (\theta)$. The mobility tensors of each Fermi surface are given as
\begin{align}
	\hat{\mu}_1 (\theta)&=\mu_0
	\begin{pmatrix}
		\beta \cos^2 \theta + \sin^2 \theta & (1-\beta)\cos \theta \sin \theta & 0
		\\
		(1-\beta) \cos \theta \sin \theta & \cos^2 \theta + \beta \sin^2 \theta & 0
		\\
		0 & 0 & 1
	\end{pmatrix},
	\\
	\hat{\mu}_2 (\theta)&=\mu_0
	\begin{pmatrix}
		\cos^2 \theta + \beta\sin^2 \theta & (\beta-1)\cos \theta \sin \theta & 0
		\\
		(\beta -1) \cos \theta \sin \theta & \beta \cos^2 \theta + \sin^2 \theta & 0
		\\
		0 & 0 & 1
	\end{pmatrix}.
\end{align}
Then, the MR is obtained as
\begin{align}
	\frac{\Delta \rho_{\perp}(\theta)}{\rho_0} &= \frac{\beta (1-\beta)^2 (\mu_0 B)^2}{(1+\beta)^2 +4\beta^2 (\mu_0 B)^2 },
	\label{TMR2}
	\\
	\frac{\Delta \rho_\parallel (\theta)}{\rho_0}&=\frac{(1-\beta)^2 (\mu_0 B)^2}{2[1+\beta + 2\beta (\mu_0 B)^2]} \sin^2 2\theta .
	\label{LMR2}
\end{align}
Unexpectedly, the longitudinal MR arises, with no change in the transverse MR. $\Delta \rho_\parallel $ is proportional to the anisotropy, $(1-\beta)^2$, and $\sin^2 2\theta$, i.e., $\Delta \rho_\parallel $ becomes largest at $\theta =\pi/4$.
\yf{
For $\theta =\pi/4$, we obtain
\begin{align}
	\frac{\Delta \rho_{\parallel}}{\Delta \rho_{\perp}}
	=\frac{(1+\beta)^2 + 4\beta^2 (\mu_0 B)^2}{2\beta (1+\beta) + 4\beta^2 (\mu_0 B)^2}.
\end{align}
At low fields, $\mu_0 B \ll 1$, $\Delta \rho_{\parallel}/\Delta \rho_{\perp}= (1+\beta)/2\beta>1$, and $\Delta \rho_{\parallel}/\Delta \rho_{\perp}$ reaches unity in the high field limit. This is a clear evidence that the longitudinal MR is always larger than the transverse MR for whole region of the magnetic field at $\theta = \pi/4$.
}
The apparent value of $\Delta \rho_\parallel$ being zero, Eq. \eqref{LMR1}, happened by accident due to $\theta =0$.
\yf{In other words, there is no reason that $\Delta \rho_{\parallel}$ is smaller than $\Delta \rho_{\perp}$ except for the angle factor, contrary to the common belief that $\Delta \rho_{\parallel}\ll \Delta \rho_{\perp}$.
}
Both the longitudinal MR and the transverse MR increase as $B^2$ at low fields ($\mu_0 B \lesssim 1$), and they saturate at high fields ($\mu_0 B \gg 1$) as shown in  Fig. \ref{Fig7} (a).

The key for the large longitudinal MR is the off-diagonal elements in $\hat{\mu}_i$. The off-diagonal elements are proportional to the anisotropy and $\sin 2\theta$, which provide the coefficient of $\Delta \rho_\parallel (\theta) $ (Eq. \eqref{LMR2}). When the direction of $\bm{J}$ (and $\bm{B}$) is along the principal axes of the mobility (or the Fermi surface), there is no off-diagonal elements in $\hat{\mu}_i$, resulting in $\Delta \rho_\parallel =0$. On the other hand, when $\bm{J}$ is diverted from the principal axes of the mobility, the off-diagonal elements in $\hat{\mu}_i$ become finite, and they generate the finite $\Delta \rho_\parallel$. The mechanism of the longitudinal MR can be understood intuitively as follows: When $\bm{J}\parallel \bm{B}$, the Lorentz force $\bm{v} \times \bm{B}$ does not work for carriers---an ordinary explanation for the absence of the longitudinal MR. This is true when $\hat{\mu}$ is diagonal. However, when the off-diagonal elements in $\hat{\mu}_i$ are finite, the velocity components are not parallel to the current as $\bm{v}=\hat{\mu}\cdot \bm{E}$. Such carriers that are not parallel to the current experience the Lorentz force, generating $\Delta \rho_\parallel$.

\begin{figure}
	\begin{center}
		\includegraphics[width=8cm]{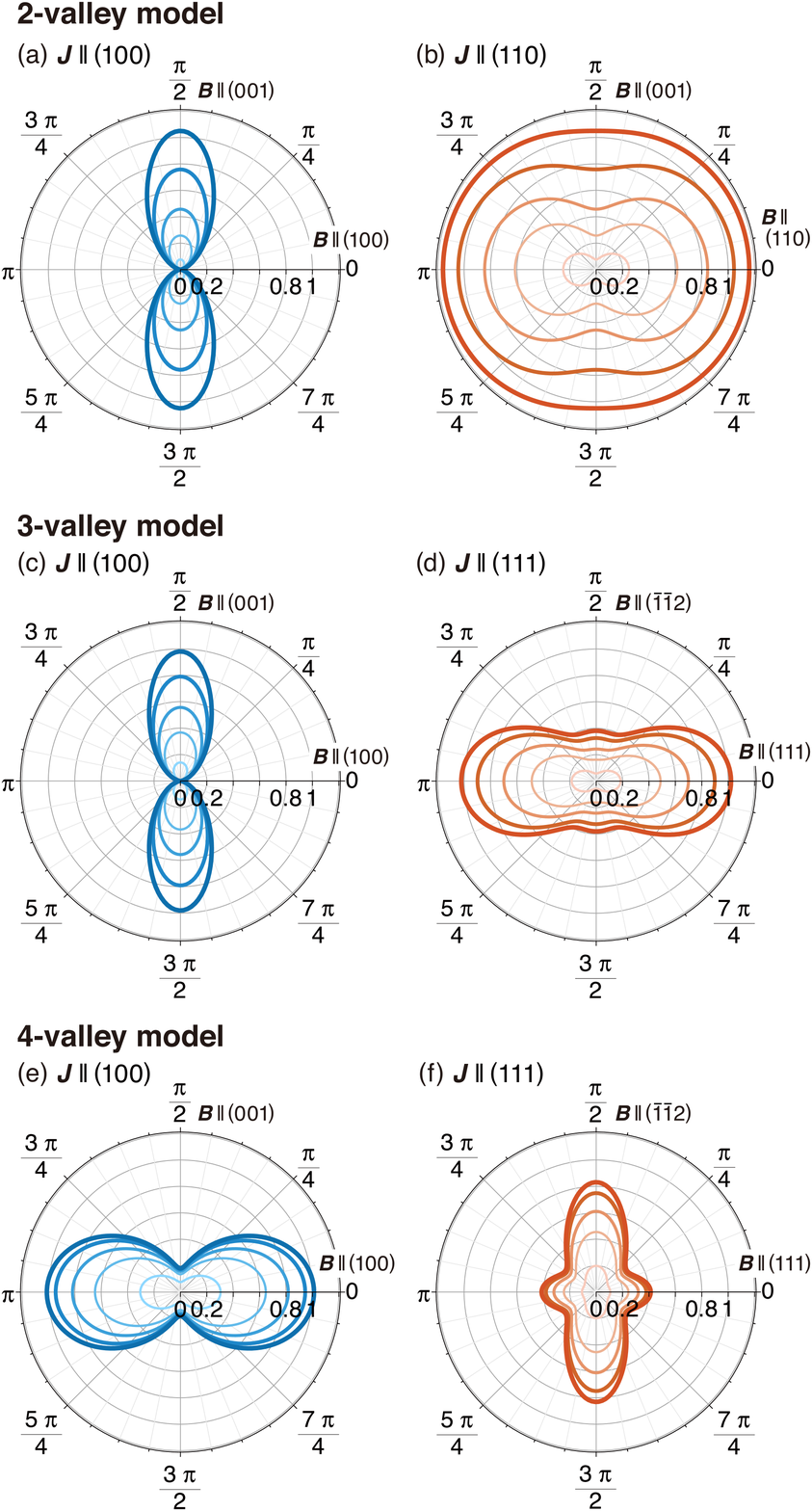}
		\caption{\label{Fig4} Polar plots of MR with respect to the rotation of the magnetic field for (a, b) two-valley ($\beta=0.15$), (c, d) three-valley ($\beta=0.15$), and (e, f) four-valley models ($\gamma=0.39$). The magnetic field orientation $\varphi =0$ corresponds to $\Delta \rho_\parallel$, and $\varphi =\pi/2$ corresponds to $\Delta \rho_\perp$ in all panels. The current orientation is fixed at $\bm{J}\parallel (100)$ in (a), (c), and (e) and at $\bm{J}\parallel (111)$ or $(110)$ in (b), (d), and (f). The magnetic field orientation is set to be $\bm{B}\parallel (100)$ for $\varphi=0$ and $\bm{B}\parallel (001)$ for $\varphi=\pi/2$ in (a), (c), and (e) and $\bm{B}\parallel (110)$ or $(111)$ for $\varphi=0$ and $\bm{B}\parallel (001)$ or $ (\bar{1}\bar{1}2)$ for $\varphi=\pi/2$ in (b), (d), and (f). The magnitude of the magnetic field is set to be $\mu_0 B = 1, 2, 3, 5$, and $10$.
		}
	\end{center}
\end{figure}

Figures \ref{Fig4} (a) and (b) shows the polar plots of MR considering the rotation of the {\it magnetic} field of the two-valley model. In Fig. \ref{Fig4} (a), $\bm{J}\parallel(100)$, and the magnetic field is rotated in the $(100)$--$(001)$ plane. In Fig. \ref{Fig4} (b), $\bm{J}\parallel(110)$, and the magnetic field is rotated in the $(110)$--$(001)$ plane. In all the panels of Fig. \ref{Fig4}, $\varphi=0$ corresponds to $\Delta \rho_\parallel$, and $\varphi=\pi/2$ corresponds to $\Delta \rho_\perp$. For the two-valley model with $\bm{J}\parallel(100)$, $\Delta \rho_\parallel$ never arises, whereas $\Delta \rho_\parallel$ arises for $\bm{J}\parallel(110)$ and  $\Delta \rho_\parallel > \Delta \rho_\perp$.

\subsection{Three-valley systems}
Here, we consider a system with three valleys, which are perpendicular to each other, as displayed in Fig. \ref{Fig1} (b). This situation is close to n-type Si or SrTiO$_3$, although the degeneracy of three valleys is slightly lifted in SrTiO$_3$ at low temperatures \cite{Mattheiss1972,Khalsa2012}. In addition to Eq. \eqref{mu12}, the third mobility tensor is given by
 \begin{align}
	\hat{\mu}_3= \mu_0 \begin{pmatrix}
		1 & 0 & 0 \\
		0 & 1 & 0\\
		0 & 0 & \beta
	\end{pmatrix}.
\end{align}
The derivations of $\Delta \rho_{\perp, \parallel}$ are the same as that for the two-valley model.
Fig. \ref{Fig4} (c) and (d) shows the polar plots of MR of the three-valley model.
For $\bm{J}\parallel (100)$ [Fig. \ref{Fig4} (c)], $\Delta \rho_\perp$ is finite but $\Delta \rho_\parallel$ is zero as seen in the two-valley system. The zero $\Delta \rho_\parallel$ can be attributed to the absence of off-diagonal components in the mobility tensor. However, when $\bm{J}\parallel (111)$ [Fig. \ref{Fig4} (d)], $\Delta \rho_\parallel$ becomes so large that $\Delta \rho_\parallel > \Delta \rho_\perp$. A large $\Delta \rho_\parallel$ arises because the current direction is diverted from each axis of the valleys.
\yf{Note that we used the following rotation matrix
\begin{align}
	\hat{R} =
  	\begin{pmatrix}
		1/\sqrt{3}&-1/\sqrt{2}&-1/\sqrt{6}\cr
  	1/\sqrt{3}&1/\sqrt{2}&-1/\sqrt{6} \cr
		1/\sqrt{3}&0&2/\sqrt{6}
	\end{pmatrix}
\end{align}
for the transformation from $\bm{J}\parallel(100)$ to $\parallel (111)$.}

\subsection{Four-valley systems}
In a cubic system such as PbTe, four ellipsoidal valleys appear along the direction of the body diagonal, as displayed in Fig. \ref{Fig1} (c). The corresponding mobility tensors are expressed as below:
\begin{align}
	\hat{\mu}_1 &=\mu_0
	\begin{pmatrix}
		1 & -\gamma & -\gamma \\
		-\gamma & 1 & -\gamma \\
		-\gamma & -\gamma & 1 \\
	\end{pmatrix},
	\quad
	\hat{\mu}_2 =\mu_0
	\begin{pmatrix}
		1 & -\gamma & \gamma \\
		-\gamma & 1 & \gamma \\
		\gamma & \gamma & 1 \\
	\end{pmatrix},
	\nonumber
	\\
	\hat{\mu}_3 &=\mu_0
	\begin{pmatrix}
		1 & \gamma & \gamma \\
		\gamma & 1 & -\gamma \\
		\gamma & -\gamma & 1 \\
	\end{pmatrix},
	\quad
	\hat{\mu}_4 =\mu_0
	\begin{pmatrix}
		1 & \gamma & -\gamma \\
		\gamma & 1 & \gamma \\
		-\gamma & \gamma & 1 \\
	\end{pmatrix}.
\end{align}
The parameter $\gamma$ expresses the anisotropy of each valley.
$\gamma$ takes the value between $0$ and $1/2$. The valley is isotropic for $\gamma=0$ and becomes a cylinder for $\gamma=1/2$.
\yf{($\gamma$ is related to $\beta$ in the two- or three-valley model through $\gamma\leftrightarrow (1-\beta)/(2+\beta)$.)}
In this case, the off-diagonal element appears everywhere in the mobility tensor, which is the origin of the large $\Delta \rho_\parallel$, as explained below.

\yf{
The conductivity tensor for $i=1$ is obtained by
\begin{subequations}
\begin{align}
	\sigma_{1xx}&=
		ne\mu_0\left(1+ D \mu_0^2B_x^2\right)/C, \\
	\sigma_{1yy}&=
		ne\mu_0 \left(1+ D \mu_0^2 B_y^2\right)/C, \\
	\sigma_{1zz}&=
		ne\mu_0 \left(1+ D \mu_0^2 B_z^2\right)/C,\\
	\sigma_{1xy}&=
		-ne\mu_0 \bigl[\gamma +\gamma\left(1+\gamma\right)\mu_0\left(B_x + B_y \right)\nonumber\\
		&+\left(1-\gamma^2\right)\mu_0B_z  -D \mu_0^2 B_xB_y\bigr]/C, \\
	\sigma_{1yz}&=
		-ne\mu_0 \bigl[\gamma +\gamma\left(1+\gamma\right)\mu_0\left( B_y + B_z \right)\nonumber\\
		&+\left(1-\gamma^2\right)\mu_0B_x -D\mu_0^2B_yB_z\bigr]/C,\\
	\sigma_{1zx}&=
		-ne\mu_0 \bigl[\gamma +\gamma\left(1+\gamma\right)\mu_0\left(B_z + B_x \right)\nonumber\\
		&+\left(1-\gamma^2\right)\mu_0B_y -D\mu_0^2B_zB_x\bigr]/C,
\end{align}
\end{subequations}
where
\begin{align}
		C&= 1 + (1-\gamma^2)\mu_0^2(B_x^2+B_y^2+B_z^2) \nonumber\\
		&+2\gamma (1+\gamma)\mu_0^2 (B_x B_y + B_y B_z + B_z B_x),
\end{align}
and $D=(1-2\gamma)(1+\gamma)^2=1-3\gamma^2 -2\gamma^3$ is the determinant of $\hat{\mu}_i/\mu_0^3$. The conductivity tensors for the other valleyss are obtained almost the same form. Only the difference is the sign of $\gamma$ and the magnetic field.
For $\bm{J}\parallel (100)$, the transverse MR [$\bm{B}=(0, 0, B)$] is obtained as
\begin{align}
	\frac{\mathit{\Delta} \rho_{\perp}}{\rho_0}
	=\frac{(1-\gamma^2)\gamma^2\mu_0^2 B^2}
	 {1 + (1-\gamma^2)^2 \mu_0^2 B^2},
	 \label{4rho1}
\end{align}
and the longitudinal MR [$\bm{B}=(B, 0, 0)$] is obtained as
\begin{align}
	\frac{\mathit{\Delta} \rho_{\parallel}}{\rho_0}
	=\frac{2(1+\gamma)\gamma^2\mu_0^2B^2}
	 {1 + D \mu_0^2 B^2}.
	 \label{4rho2}
\end{align}
The ratio of $\Delta \rho_{\parallel}$ to $\Delta \rho_{\perp}$ is
\begin{align}
	\frac{\Delta \rho_{\parallel}}{\Delta \rho_{\perp}}
	=\frac{2\left[ 1+(1-\gamma^2)^2 \mu_0^2 B^2\right]}{(1-\gamma)(1+D\mu_0^2 B^2)}.
	\label{4rho3}
\end{align}
At low fields, $\Delta \rho_{\parallel}/\Delta \rho_{\perp}=2/(1-\gamma)>2$, and $\Delta \rho_{\parallel}/\Delta \rho_{\perp}= 2(1-\gamma)/(1-2\gamma)>2$ at high fields. The longitudinal MR is more than twice larger than the transverse MR for whole region of the magnetic field, irrespective of the anisotropy. It is clear from Eqs. \eqref{4rho1}-\eqref{4rho3}, there is no reason that $\Delta \rho_\parallel$ is smaller than $\Delta \rho_\perp$ except for a particular angle. The common belief of $\Delta \rho_{\parallel} \ll \Delta \rho_{\perp}$ is valid only when the current direction is along the axis of valleys. This is one of the main results in the present paper.
}

Figure \ref{Fig7} (c) shows $\Delta \rho_{\perp, \parallel}$ of the four-valley model as a function of $\mu_0 B$ ($= \omega_c \tau$).
\yf{In Fig. \ref{Fig7} (c), we set $\gamma=0.39$ to be consistent with the expereiments on PbTe as discussed later. }
The solid lines are for $\bm{J}\parallel (100)$, and the dashed lines are for $\bm{J}\parallel (111)$. A notable $\Delta \rho_\parallel$ arises for $\bm{J}\parallel (100)$ . Both the longitudinal MR and the transverse MR increase as $B^2$ in the low field $\mu_0 B \lesssim 1$, and saturate at high fields $\mu_0 B \gg 1$. The saturated value of $\Delta \rho_\parallel$ for $\bm{J}\parallel (100)$ is more than 5 times larger than that of $\Delta \rho_\perp$. The large $\Delta \rho_\parallel$ can be seen more clearly if we look at the polar plots of MR [Fig. \ref{Fig4} (e) and (f)]. For $\bm{J}\parallel (100)$, $\Delta \rho_\parallel$ ($\varphi=0$) is larger than $\Delta \rho_\perp$ ($\varphi=\pi/2$).
The origin of the large $\Delta \rho_\parallel$ is the same as that for the two- and three-valley systems, but it is more highlighted here. As the axes of the four valleys are diverted from $\bm{J} \parallel (100)$, the off-diagonal elements appear everywhere in the four-valley system, so that $\Delta \rho_\parallel$ becomes larger than $\Delta \rho_\perp.$
 When the current direction is along one of the body diagonals, $\bm{J}\parallel (111)$, $\Delta \rho_\parallel$ is weakened [Fig. \ref{Fig4} (f)], because the mobility tensor of one of the valleys becomes diagonal (the other three valleys are still diverted from the current direction, so that they contribute to $\Delta \rho_\parallel$.)
Although the polar plots of MR for the four-valley system [Fig. \ref{Fig4} (e), (f)] seem to be completely opposite from that for the three-valley system [Fig. \ref{Fig4} (c), (d)], whole polar plots can be explained consistently from the universal viewpoint of off-diagonal mobility.

\begin{figure}
	\begin{center}
		\includegraphics[width=7cm]{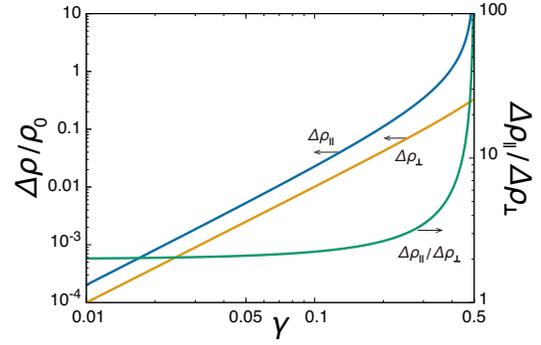}
		\caption{\label{Fig6} Anisotropy dependence of MR $\Delta \rho_{\perp, \parallel}/\rho_0$ and their ratio $\Delta \rho_{\parallel}/\Delta \rho_{\perp}$ at high fields $\mu_0 B \gg 1$ for the four-valley systems. The current direction is fixed at $\bm{J}\parallel (1 0 0 )$.
		}
	\end{center}
\end{figure}

Another important factor of the large $\Delta \rho_\parallel$ is the anisotropy of the mobility tensor, which is characterized by $\gamma$ in the current model. Figure \ref{Fig6} shows the $\gamma$-dependence of $\Delta \rho_{\perp, \parallel}$ for $\bm{J}\parallel (100)$ and the $\gamma$-dependence of their ratio $\Delta \rho_\parallel/\Delta \rho_\perp$. Both $\Delta \rho_{\perp, \parallel}$ increase as the anisotropy increases. Furthermore, the ratio $\Delta \rho_\parallel/\Delta \rho_\perp$ also increases as $\gamma$ increases, and it finally diverges at $\gamma = 1/2$.

\section{Comparison with experiments}
\begin{figure}
	\begin{center}
		\includegraphics[width=7cm]{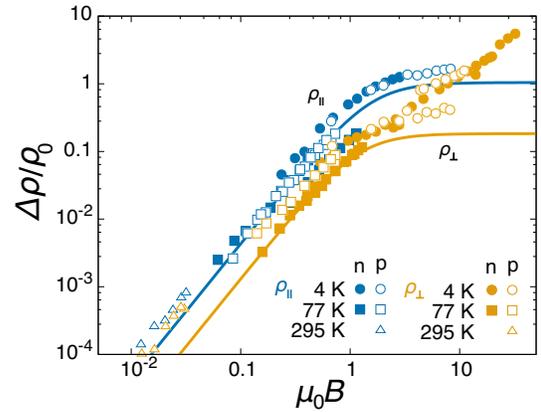}
		\caption{\label{Fig5} Magnetoresistance of experiments (symbols) by Allgaier \cite{Allgaier1958} and the present theory (solid lines) of the four-valley model with $\gamma=0.39$ as a function of $\mu_0 B$ ($=\omega_c \tau$). Note that here we plot the results in the SI unit instead of the CGS unit, which was used in the original work by Allgaier \cite{Allgaier1958}.}
	\end{center}
\end{figure}

Lastly, we compare the theoretical results with the experimental results. We summarize the experimental data of PbTe by Allgaier for five samples, including n-type and p-type carriers at $T=$ 4.2, 77.4, and 295 K [Fig. \ref{Fig5}]. The experimental data are scaled into a single curve as a function of $\mu_0 B$ ($=\omega_c \tau$) for each $\Delta \rho_\perp$ and $\Delta \rho_\parallel$.
\yf{(Allgaier already scaled the experimental data as a function of $\mu_0 B$, but plotted them in separate figures for each sample. Here we replot them into a single figure.)}
This scaling property signifies that the following properties are intrinsic to MR in PbTe: (i) $\Delta \rho_{\perp, \parallel}\propto B^2$ for $\mu_0 B \lesssim 1$, and (ii) $\Delta \rho_{\perp, \parallel} \propto B^0$ for $\mu_0 B \gg 1$. The only exception is $\Delta \rho_\perp$ in the large $\mu_0 B$ region at 4K. There are two behaviors: $\Delta \rho_\perp$ is saturated at $\mu_0 B \gg 1$ in two samples (both n-type and p-type), and it increases linearly and is unsaturated at $\mu_0 B \gg 1$ in one sample.

The theoretical lines in Fig. \ref{Fig5} denote the results with $\gamma = 0.39$, \yf{which is determined so as to fit the experimental data. ($\gamma=0.39$ corresponds to $\beta=0.15$). }This anisotropy is consistent with the band calculation of PbTe \cite{Lent1986,Izaki_private}. It is unexpected that both $\Delta \rho_{\perp}$ and $\Delta \rho_\parallel$ can be well fitted simultaneously into a single parameter $\gamma$. Therefore, we conclude that the intrinsic property of MR in PbTe can be quantitatively explained only within the classical theory of MR, even at high fields $\mu_0 B \gtrsim 1$ for a wide range of temperatures---4 K to room temperature.

Allgaier also reported the MR of PbS and PbSe. The data are not sufficient to check the scaling behavior of each $\Delta \rho_{\perp, \parallel}$. However, we have concluded two properties of MR in PbS and PbSe: the magnitudes of $\Delta \rho/\rho_0$ and the ratios of $\Delta \rho_{\parallel}/\Delta \rho_{\perp}$ in PbS and PbSe are smaller than those of PbTe. This is consistent with the observations of Kawakatsu {\it et al.} \cite{Kawakatsu2019}.
These properties can be explained in terms of anisotropy. The Fermi surfaces of PbS and PbSe are more isotropic than that of PbTe \cite{Krizman2018,Kawakatsu2019}. According to our results, $\Delta \rho_{\perp, \parallel}/\rho_0$ increases as $\gamma$ increases [Fig. \ref{Fig6}].

The linear MR in PbTe is an extrinsic property, but it appears only when  $\mu_0 B >1$; i.e., it can be a characteristic of high mobility samples. The linear MR in PbTe has been observed repeatedly by \cite{Shogenji1959,Kawakatsu2019}. The sample used by Kawakatsu {\it et al.} exhibits the linear $\Delta \rho_\perp$ around 1 T. This would be because of the high purity of the sample with $\mu_0$=31 T$^{-1}$ ($\rho_0 = 9.3$ $\mu \Omega$cm), which sufficiently satisfies the condition $\mu_0 B\gg 1$. According to our results, the linear MR can never be explained within the classical theory, although the MR can be apparently linear in the intermediate region $\mu_0 B \sim 1$.

\section{Conclusions}
We have investigated the MR in the multivalley systems based on the classical theory. It has been clarified that the notable longitudinal MR, $\Delta \rho_\parallel$, is realized when the current direction is diverted from the axis of the ellipsoidal Fermi surface of each valley, which sets aside the common belief.
\yf{Analytical proofs that $\Delta \rho_\parallel$ is larger than $\Delta \rho_\perp$ are given for a certain direction of the current off from the axis of the ellipsoids.}
The large $\Delta \rho_\parallel$ originates from the off-diagonal components of the mobility tensor. The off-diagonal components transfer the velocity of electrons from parallel to non-parallel toward the magnetic field direction, so that the Lorentz force works on electrons, generating the longitudinal MR.
\yf{This is the first time to give a clear relationship between the large longitudinal MR, $\Delta \rho_{\parallel} > \Delta \rho_{\perp}$, and the axis of the ellipsoids (the off-diagonal mobility tensor), which is quite helpful for the intuitive prediction.}

Our main finding is that there is no reason that $\Delta \rho_\parallel$ should be smaller than $\Delta \rho_\perp$. It depends on the current orientation. $\Delta \rho_\parallel$ can exceed $\Delta \rho_\perp$ when the current is diverted from the axes of all the valleys.
\yf{The common belief of $\Delta \rho_{\parallel} \ll \Delta \rho_{\perp}$ is valid only when the current direction is along the axis of valleys.}
In the case of the four-valley system, where the axis of each ellipsoidal Fermi surface is oriented along the four body-diagonal directions of the crystal, $\Delta \rho_\parallel$ is larger than $\Delta \rho_\perp$, even at high fields $\mu_0 B \gg 1$.

Our classical theory can very well agree with the experiments on PbTe, where $\Delta \rho_\parallel$ is larger than $\Delta \rho_\perp$ at 4 K to room temperature with n- and p-type samples. Only one exception is the linear transverse MR at 4 K, which can never be explained based on the present classical theory (and the semi-classical theory using the Boltzmann equation; our classical formula is essentially the same as the semi-classical formula). Recently, it was shown that the results obtained by the classical formula (and the semi-classical formula) almost agreed with those obtained by the quantum theory based on the Kubo formula assuming the constant relaxation time $\tau$ \cite{Owada2018}. Therefore, the quantum theory with constant relaxation time will not be able to explain the linear MR as well. The linear MR may be explained by the extrinsic mechanism, i.e., the field dependence of the relaxation time \cite{ZZhu2018,Fauque2019}, which will be discussed in future studies.

\section*{Acknowledgments}
We would like to thank Y. Onuki, M. Tokunaga, K. Akiba, K. Behnia, and B. Fauqu\'e for their helpful discussions. This work is supported by JSPS KAKENHI (Grant No. 19H01850 and Grant No. 16K05437).

\section*{References}
\providecommand{\newblock}{}


\begin{thebibliography}{10}
\expandafter\ifx\csname url\endcsname\relax
  \def\url#1{{\tt #1}}\fi
\expandafter\ifx\csname urlprefix\endcsname\relax\def\urlprefix{URL }\fi
\providecommand{\eprint}[2][]{\url{#2}}

\bibitem{Thomson1857}
Thomson W 1857 {\em Proc. R. Soc. Lond.\/} {\bf 8} 546

\bibitem{Ziman_book}
Ziman J~M 1972 {\em Principles of the Theory of Solids\/} 2nd ed (Cambridge
  University Press)

\bibitem{Kittel_book_quantum}
Kittel C 1987 {\em Quantum Theory of Solids\/} (Wiley) ISBN 9780471624127

\bibitem{Abrikosov_book}
Abrikosov A 1988 {\em Fundamentals of the Theory of Metals\/} (North-Holland)
  ISBN 9780444870940

\bibitem{Beer1963}
Beer A 1963 {\em Galvanomagnetic effects in semiconductors\/} Solid State
  Physics Series (Academic Press)

\bibitem{Pippard_book}
Pippard A 1989 {\em Magnetoresistance in Metals\/} Cambridge Studies in Low
  Temperature Physics (Cambridge University Press) ISBN 9780521326605

\bibitem{Kapitza1928}
Kapitza P 1928 {\em Proc. Roy. Soc. A\/} {\bf 119} 358

\bibitem{Taub1971}
Taub H, Schmidt R~L, Maxfield B~W and Bowers R 1971 {\em Phys. Rev. B\/} {\bf
  4}(4) 1134--1152

\bibitem{RXu1997}
Xu R, Husmann A, Rosenbaum T~F, Saboungi M~L, Enderby J~E and Littlewood P~B
  1997 {\em Nature\/} {\bf 390} 57--60

\bibitem{Narayanan2015}
Narayanan A, Watson M~D, Blake S~F, Bruyant N, Drigo L, Chen Y~L, Prabhakaran
  D, Yan B, Felser C, Kong T, Canfield P~C and Coldea A~I 2015 {\em Phys. Rev.
  Lett.\/} {\bf 114}(11) 117201

\bibitem{ZZhu2018}
Zhu Z, Fauqu{\'e} B, Behnia K and Fuseya Y 2018 {\em J. Phys.: Condens.
  Matter\/} {\bf 30} 313001

\bibitem{Kawakatsu2019}
Kawakatsu S, Nakaima K, Kakihana M, Yamakawa Y, Miyazato H, Kida T, Tahara T,
  Hagiwara M, Takeuchi T, Aoki D, Nakamura A, Tatetsu Y, Maehira T, Hedo M,
  Nakama T and {\=O}nuki Y 2019 {\em J. Phys. Soc. Jpn.\/} {\bf 88} 013704

\bibitem{Arnold2016}
Arnold F, Shekhar C, Wu S~C, Sun Y, dos Reis R~D, Kumar N, Naumann M, Ajeesh
  M~O, Schmidt M, Grushin A~G, Bardarson J~H, Baenitz M, Sokolov D, Borrmann H,
  Nicklas M, Felser C, Hassinger E and Yan B 2016 {\em Nature Communications\/}
  {\bf 7} 11615

\bibitem{JXu2019}
Xu J, Ma M~K, Sultanov M, Xiao Z~L, Wang Y~L, Jin D, Lyu Y~Y, Zhang W, Pfeiffer
  L~N, West K~W, Baldwin K~W, Shayegan M and Kwok W~K 2019 {\em Nature
  Communications\/} {\bf 10} 287

\bibitem{Allgaier1958}
Allgaier R~S 1958 {\em Phys. Rev.\/} {\bf 112}(3) 828--836

\bibitem{Gupta1978}
Gupta S~C, Rajwanshi K~N~S and Sreedhar A~K 1978 {\em Journal of Applied
  Physics\/} {\bf 49} 469--470

\bibitem{Onuki_private}
Onuki Y 2018 (private communication)

\bibitem{Abeles1954}
Abeles B and Meiboom S 1954 {\em Phys. Rev.\/} {\bf 95}(1) 31--37

\bibitem{Shibuya1954}
Shibuya M 1954 {\em Phys. Rev.\/} {\bf 95}(6) 1385--1393

\bibitem{Gold1957}
Gold L and Roth L~M 1957 {\em Phys. Rev.\/} {\bf 107}(2) 358--364

\bibitem{Roth1992}
Roth L~M 1992 {\em Dynamics and classical transport of carriers in
  semiconductors\/} (North Holland) chap~10, p 489

\bibitem{Askerov1994}
Askerov B~M 1994 {\em Electron Transport Phenomena in Semiconductors\/} (WORLD
  SCIENTIFIC)

\bibitem{Mackey1969}
Mackey H~J and Sybert J~R 1969 {\em Phys. Rev.\/} {\bf 180}(3) 678--681

\bibitem{Awashima2019}
Awashima Y and Fuseya Y 2019 {\em Journal of Physics: Condensed Matter\/} {\bf
  31} 29LT01

\bibitem{Aubrey1971}
Aubrey J~E 1971 {\em Journal of Physics F: Metal Physics\/} {\bf 1} 493

\bibitem{Collaudin2015}
Collaudin A, Fauqu\'e B, Fuseya Y, Kang W and Behnia K 2015 {\em Phys. Rev.
  X\/} {\bf 5}(2) 021022

\bibitem{Mattheiss1972}
Mattheiss L~F 1972 {\em Phys. Rev. B\/} {\bf 6}(12) 4718--4740

\bibitem{Khalsa2012}
Khalsa G and MacDonald A~H 2012 {\em Phys. Rev. B\/} {\bf 86}(12) 125121

\bibitem{Lent1986}
Lent C~S, Bowen M~A, Dow J~D and Allgaier R~S 1986 {\em Superlattices
  Microstruct.\/} {\bf 2} 491

\bibitem{Izaki_private}
Izaki Y 2018 (private communication)

\bibitem{Krizman2018}
Krizman G, Assaf B~A, Phuphachong T, Bauer G, Springholz G, de~Vaulchier L~A
  and Guldner Y 2018 {\em Phys. Rev. B\/} {\bf 98}(24) 245202

\bibitem{Shogenji1959}
Shogenji K 1959 {\em Journal of the Physical Society of Japan\/} {\bf 14}
  1360--1371

\bibitem{Owada2018}
Owada M, Awashima Y and Fuseya Y 2018 {\em Journal of Physics: Condensed
  Matter\/} {\bf 30} 445601

\bibitem{Fauque2019}
Collignon C, et al., to be submitted.

\end{thebibliography}

\end{document}